\documentclass[conference]{IEEEtran}
\IEEEoverridecommandlockouts
\usepackage{cite}
\usepackage{amsmath,amssymb,amsfonts}
\usepackage{algorithmic}
\usepackage{graphicx}
\usepackage{textcomp}
\usepackage[dvipsnames]{xcolor}
\usepackage{soul}
\definecolor{myblue}{RGB}{0, 114, 178}
\usepackage[colorlinks=true, allcolors=myblue]{hyperref}
\def\BibTeX{{\rm B\kern-.05em{\sc i\kern-.025em b}\kern-.08em
    T\kern-.1667em\lower.7ex\hbox{E}\kern-.125emX}}
\begin{document}

\title{Erasure Minesweeper: exploring hybrid-erasure surface code architectures for efficient\\quantum error correction
\thanks{

\noindent\href{mailto:jchadwick@uchicago.edu}{$^*$jchadwick@uchicago.edu}\\
\href{mailto:mhteo@uchicago.edu}{$^*$mhteo@uchicago.edu}\\
\href{mailto:viszlai@uchicago.edu}{$^*$viszlai@uchicago.edu}\\
\href{mailto:willers@uchicago.edu}{$^*$willers@uchicago.edu}
}
}

\author{\IEEEauthorblockN{Jason D. Chadwick$^*$, Mariesa H. Teo$^*$, Joshua Viszlai$^*$, Willers Yang$^*$, and Frederic T. Chong}
\IEEEauthorblockA{\textit{Department of Computer Science, University of Chicago} \\
Chicago, IL, USA \\
$^*$Authors contributed equally; listed alphabetically
}}
\maketitle
\thispagestyle{plain}
\pagestyle{plain}

\begin{abstract}

Dual-rail erasure qubits can substantially improve the efficiency of quantum error correction, allowing lower error rates to be achieved with fewer qubits, but each erasure qubit requires $3\times$ more transmons to implement compared to standard qubits. In this work, we introduce a hybrid-erasure architecture for surface code error correction where a carefully chosen subset of qubits is designated as erasure qubits while the rest remain standard. Through code-capacity analysis and circuit-level simulations, we show that a hybrid-erasure architecture can boost the performance of the surface code---much like how a game of Minesweeper becomes easier once a few squares are revealed---while using fewer resources than a full-erasure architecture. We study strategies for the allocation and placement of erasure qubits through analysis and simulations. We then use the hybrid-erasure architecture to explore the trade-offs between per-qubit cost and key logical performance metrics such as threshold and effective distance in surface code error correction. Our results show that the strategic introduction of dual-rail erasure qubits in a transmon architecture can enhance the logical performance of surface codes for a fixed transmon budget, particularly for near-term-relevant transmon counts and logical error rates.\end{abstract}


\newcommand{\todocite}[0]{{\color{red}[CITE]}}

\section{Introduction}

\begin{figure}
    \centering
    \includegraphics[width=0.95\linewidth]{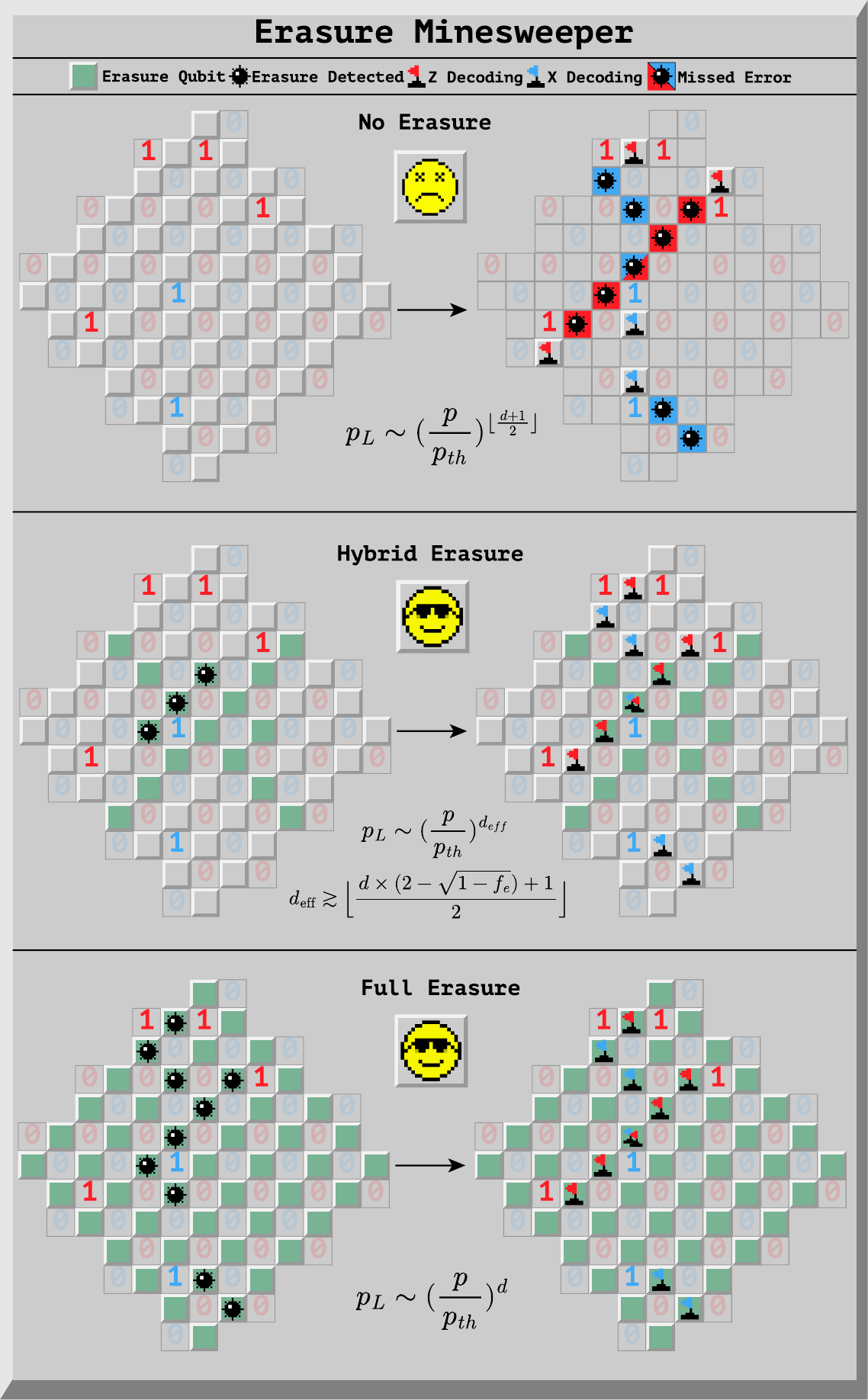}
    \caption{Viewing a hybrid-erasure architecture as a game of Minesweeper. \emph{Top:} In a standard surface code, we receive syndrome bits (red and blue 1s) indicating that an error string terminates nearby. The decoder must find a valid explanation (red and blue flags) based only on this information. If the decoder guesses wrong, a logical error can occur. \emph{Bottom:} When data qubits are \emph{erasure qubits}, we gain extra information before decoding (black bombs and green checks on left) that help us decide which error string to predict, avoiding logical errors and achieving better error suppression. \emph{Middle:} Partial improvements over error suppression persists when a carefully selected a subset of qubits are erasure qubits. }
    \label{fig:minesweeper-overview}
\end{figure}

Quantum error correction (QEC) is expected to be necessary to achieve the low error rates of $10^{-12}$ to $10^{-18}$ needed to enable powerful applications of quantum computing \cite{litinski_game_2019, gidney_how_2021, beverland_assessing_2022, chamberland_universal_2022, kim_fault-tolerant_2022, leblond_realistic_2024}. QEC offers the promise of exponential error suppression with increasing code size once physical qubit errors are below a certain \emph{threshold}. The surface code is a leading approach to quantum error correction due to its relatively high threshold, compatibility with planar nearest-neighbor qubit connectivity, and ease of decoding \cite{fowler_surface_2012}.

While typical noise models used to simulate and evaluate QEC assume that physical one- and two-qubit errors are random, uniformly distributed Pauli operators, significant gains in QEC efficiency can be found under alternative noise models \cite{bonilla_ataides_xzzx_2021, dua_clifford-deformed_2024, tuckett_ultrahigh_2018, tuckett_tailoring_2019, gu_fault-tolerant_2023, sahay_high-threshold_2023}. For example, under a biased noise model where the relative strengths of Pauli X and Z errors are orders of magnitude different, the XZZX surface code can drastically improve the noise threshold compared to the standard surface code \cite{bonilla_ataides_xzzx_2021}. Another type of non-standard error model is \emph{heralded erasure}, in which the information in a physical qubit is completely lost, but the time and location of the error are known. A surface code made up of \emph{erasure qubits}, in which the primary error mechanism is heralded erasure, can have a higher threshold and, crucially, twice the effective distance: a surface code patch with $d \times d$ data qubits can correct $d-1$ heralded erasure errors on ideal erasure qubits, but only $\lfloor (d-1)/2 \rfloor$ errors under a standard Pauli error model. This gives erasure qubits a significant scaling advantage over standard unbiased qubits, leading to significant interest in the proposal \cite{wu_erasure_2022, kubica_erasure_2023, teoh_dual-rail_2023, kang_quantum_2023} and realization \cite{ma_high-fidelity_2023, scholl_erasure_2023, chou_demonstrating_2023, koottandavida_erasure_2023, levine_demonstrating_2024, chow_circuit-based_2024, holland_demonstration_2024, de_graaf_mid-circuit_2024, mehta_bias-preserving_2025} of erasure qubits in various hardware modalities.

\emph{Dual-rail} erasure qubits in superconducting transmons \cite{teoh_dual-rail_2023, kubica_erasure_2023} are a particularly promising approach to implementing erasure qubits, with several recent experimental realizations \cite{levine_demonstrating_2024, chou_demonstrating_2023, mehta_bias-preserving_2025}. A dual-rail qubit is encoded in two transmons such that a relaxation error (the dominant error in individual transmons) in either transmon takes the system out of the qubit space; an \emph{erasure check} can be performed regularly to determine whether this has happened, and if so, the qubit is reset and the information is sent to the decoder. Some dual-rail implementations require a third transmon to perform this erasure check, bringing the relative cost of a single dual-rail erasure qubit to $3\times$ that of a standard qubit.

There is therefore a trade-off between increased complexity of each physical qubit and improved QEC efficiency. Several factors may impose limits on the number of transmons that can be used for each logical qubit in the surface code. First, fabrication defects are a considerable hurdle for large-scale transmon-based quantum computers, as the \emph{yield} (proportion of fully functional chips without defects) decreases with increasing transmon count per chip. The defect rate in current industrial devices is proprietary, but is commonly estimated to be between $10^{-3}$ and $2 \times 10^{-2}$ \cite{kreikebaum_improving_2020, zeissler_superconducting_2024, debroy_luci_2024}. Although modular approaches are expected to mitigate this problem \cite{smith_scaling_2022}, it may be desirable for each logical qubit to be fully contained on a single chip if chip-to-chip connections are restricted \cite{lin_codesign_2024}. Second, once yield is addressed, there will still be limits to the maximum number of transmons that can be placed into a single dilution refrigerator due to heating concerns \cite{hashimoto_demonstration_2005, krinner_engineering_2019, byun_xqsim_2022, ueno_inter-temperature_2024}; larger transmon counts will require distributed multi-fridge approaches that will have reduced inter-fridge communication capabilities \cite{jnane_multicore_2022}.

In this work, we investigate a hybrid-erasure architecture for implementing surface codes in transmons, where physical qubits of the surface code are composed of a mixture of standard and dual-rail erasure qubits. As key benefits of erasure come from the additional error information, optimizing such a hybrid-erasure architecture rests on the allocation of erasure qubits---deciding how many erasure qubits to use and where to place them---so the additional information can best be exploited to reach a target logical error performance. Erasure qubits supply the decoder with additional information on where the errors have occurred (and more importantly, have \emph{not} occurred), allowing the decoder to more efficiently distinguish topologically distinct error chains. The objective is similar to that of Minesweeper, where tiles are revealed strategically to deduce the locations of bombs (physical errors). To this end, we first study the surface code in a code-capacity model, allowing us to determine an optimized placement strategy. Then, through more accurate circuit-level simulations, we evaluate the QEC performance of a hybrid-erasure surface code using this placement heuristic and study the trade-offs between key metrics such as effective distance, threshold, and logical error rates. We find that our placement heuristic results in a hybrid-erasure chip that achieves better logical error rate than a homogeneous all-standard or all-erasure qubit chips for certain relevant transmon-count budgets.


The remainder of the paper is structured as follows. In Section~\ref{sec:background}, we introduce erasure qubits and surface code error correction. Thus far, studies of erasure for error correction have assumed a homogeneous, all-erasure layout. We extend the prior work by introducing a hybrid-erasure architecture for surface code error correction in Section~\ref{sec:hybrid-arch}, and analyze, in a code-capacity model, the impact of design choices such as allocation and placement on the logical performance. In Section \ref{sec:eval-method}, we describe our circuit-level simulation framework. In Section~\ref{sec:eval-result}, we discuss the results of these circuit-level evaluations and analyze transmon costs, chip yield, and logical performance for varying system sizes using a hybrid-erasure chip versus homogeneous all-standard or all-erasure chips. We also explore alternative placement strategies in Section~\ref{sec:empirical} under circuit-level noise and compare simulation results to the predicted behavior from the code-capacity analysis. Finally, in Section~\ref{sec:discussion}, we discuss conclusions and future work.

\begin{figure}
    \centering
    \includegraphics[width=0.6\linewidth]{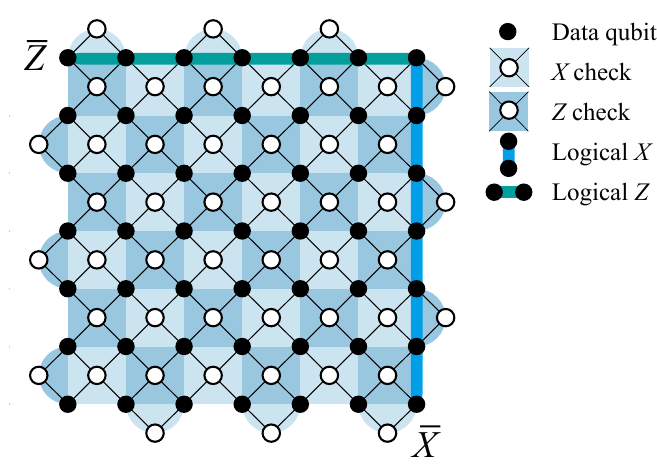}
    \caption{A $d=7$ rotated surface code patch, consisting of $d^2=49$ data qubits (black) and $d^2-1=48$ ancilla qubits (white) in a degree-four planar grid. X and Z stabilizers are represented by light and dark plaquettes, indicating weight-two or weight-four parity checks. The $\bar X$ and $\bar Z$ logical operators are highlighted in light and dark blue lines at the boundary. In the ``Erasure Minesweeper'' game of Figure \ref{fig:minesweeper-overview}, data qubits are potential bomb tiles and ancilla qubits are tiles with numbers.}
    \label{fig:surface-code}
\end{figure}

\begin{figure*}[t]
    \centering
    \includegraphics[width=0.9\linewidth]{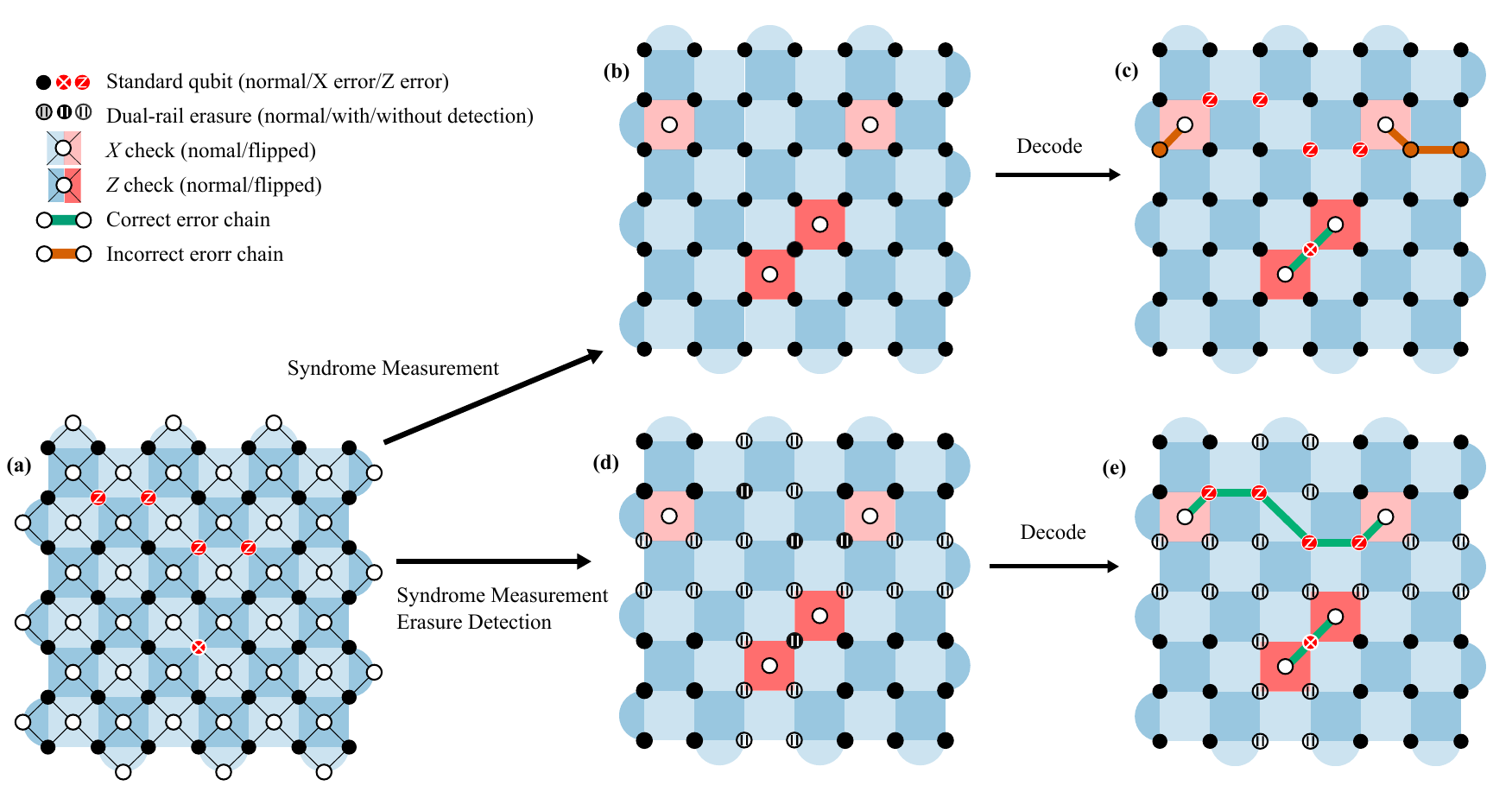}
    \caption{Decoding an example set of errors (a) in the surface code. (b) Syndrome information available in a standard, erasure-free surface code. (c) Decoder prediction resulting in a logical Z error. (d) Syndrome + erasure information available in a hybrid erasure architecture with two rows/columns of erasure qubits. (e) Decoder prediction resulting in no logical errors. The lack of erasure detection on the erasure qubits near the boundary critically help the decoder differentiate the correct error chain.}
    \label{fig:decoding-surface-code}
\end{figure*}

\section{Background}\label{sec:background}

\subsection{The surface code}


A distance-$d$ surface code (Figure~\ref{fig:surface-code}) encodes one logical qubit in the entangled state of $d^2$ physical qubits on a square grid. The logical states correspond to the eigenstates of a commuting group of $d^2-1$ $X$-type and $Z$-type parity check operators. These check operators have weight $2$ or $4$, meaning they're supported on $2$ or $4$ physical qubits, and can be measured using only 2D local interactions with measurement ancillae interleaved with the data qubits. The logical observables $\bar X$ and  $\bar Z$  
 of the logical qubit are encoded non-locally as the parity of physical $X$s and $Z$s on a path crossing the lattice, the minimum lengths of which grow as we increase the size of the lattice.

A physical error can be detected by periodically measuring the parity check operators and corrected if a decoder can determine a valid set of physical errors that match the measurement outcomes. The outcome of the parity check operators is often referred to as the error syndrome. In the absence of physical errors, the syndrome is invariant; and when a chain of physical errors occurs, the checks on the ends of the error chain will be flipped, yielding a non-trivial syndrome. The decoder attempts to correct the error by finding the most likely explanation given the error syndrome. Due to the non-local encoding of the logical observables as a lattice-traversing path, an error chain can cause a logical error only if it traverses the lattice, which can occur if the decoder guesses incorrectly and applies an invalid correction. We denote the probability of a single physical error as $p$ and the probability of a logical error as $p_L$.

The relationship between $p$ and $p_L$ of an error-correcting code depends critically on two metrics: the threshold $p_{th}$, which characterizes the minimum physical error rate $p$ required to achieve suppression of $p_L$ with increasing code size, and the effective distance $d_{\text{eff}}$, which characterizes how sensitive $p_L$ is to $p$.

\subsection{Erasure qubits}

An erasure qubit is a physical qubit whose primary error mechanism is a heralded erasure, which is a depolarizing error that occurs at a known time. A surface code implemented using ideal erasure qubits is known to experience large improvements in threshold and effective distance \cite{stace_thresholds_2009, delfosse_almost-linear_2021, sahay_high-threshold_2023}. In practice, the degree of this scaling advantage depends on the physical implementation of the erasure qubit \cite{sahay_high-threshold_2023, gu_fault-tolerant_2023}. Here, we briefly review several prominent proposals for erasure qubit implementations, designed to convert the dominant errors in the hardware into erasure errors. 

Dual rail transmon qubits, which are the focus of this work, comprise two superconducting transmons coupled together \cite{kubica_erasure_2023}. The logical computational basis states are given by 
\begin{equation}
    |\overline{0}\rangle = \frac{|01\rangle + |10\rangle}{\sqrt{2}} \hspace{6pt},\hspace{18pt} |\overline{1}\rangle = \frac{|01\rangle - |10\rangle}{\sqrt{2}} 
\end{equation}

In each transmon, amplitude damping errors dominate, which bring the dual rail erasure qubit to the $|00\rangle$ state. If an erasure qubit is detected to be in this state, it can be reset, returning it to the computational basis \cite{gu_optimizing_2024, magnard_fast_2018, marques_all-microwave_2023, zhou_rapid_2021, geerlings_demonstrating_2013}. In total, this implementation requires two transmons to make up the dual-rail, and an additional one for the erasure check, bringing its total physical qubit cost to 3 per erasure qubit. We note that some proposals have suggested using a superconducting cavity for readout \cite{kubica_erasure_2023}, but for simplicity here our resource estimates are quoted in terms of the three-transmon implementation. 

Beyond transmons, erasure qubit implementations have been proposed and demonstrated in dual-rail superconducting cavities, where dominant photon loss errors are detected and converted into erasures\cite{koottandavida_erasure_2023, de_graaf_mid-circuit_2024, teoh_dual-rail_2023, chou_demonstrating_2023}. There has also been significant effort in both theory and experiment in developing erasure qubits in neutral atoms, and it is estimated that up to 98\% of leakage outside the qubit subspace can be converted into erasure \cite{wu_erasure_2022, ma_high-fidelity_2023}. A similar scheme has been suggested for erasure conversion in trapped ion systems \cite{kang_quantum_2023}. We focus on dual-rail qubits in this work because they are fabricated from the same components as standard superconducting qubits, so we can make a direct cost comparison between them. However, much of our findings can also be generalized to other hybrid erasure schemes.

\section{Analysis of a hybrid-erasure surface code}\label{sec:hybrid-arch}

We consider surface code implemented on a hybrid-erasure architecture $\mathcal{A}(d,f_e,P)$, where
\begin{itemize}
    \item $d\in  \{3,5,..\}$ is the surface code distance,
    \item $f_e\in[0,1]$ upper-bounds the fraction of erasure qubits, 
    \item $P\subset[d^2]$ specifies the subset of data qubits (indexed by $[d^2] = \{1,..,d^2\}$) designated as erasures.
\end{itemize}
Note that the two edge cases, $\mathcal{A}(d,0,0)$ and $\mathcal{A}(d,1,[d^2])$, correspond to a distance-$d$ surface code on all standard qubits and all erasure qubits, respectively.

We perform analysis in the code-capacity model, in which errors can occur on data qubits but syndrome measurements are assumed to be perfect (and there is no propagation of error between qubits). Section~\ref{ssec:simulation-validation} confirms that our analysis closely matches circuit‑level simulation results.

\subsection{Summary of results}


\begin{figure}
    \centering
    \includegraphics[width=0.6\linewidth]{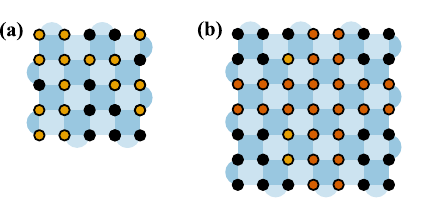}
        \caption{Examples of hybrid-erasure architecture. (a) $\mathcal{A}(5,0.6,P_r)$ where the placement $P_r = \{1,2,5,...,22\}$ is sampled at random. (b) $\mathcal{A}(7,0.57,P^*)$. Since $f_e\times d^2 = 27.93$, only $27$ erasures are allocated, with $24$ erasures placed in $2$ rows/columns, and the remaining $3$ placed greedily, all as close to the center as possible.}
    \label{fig:placements}
\end{figure}

We analyze the performance of surface codes on a hybrid-erasure architecture $\mathcal{A}(d, f_e,P^*)$ in a code-capacity model, and prove that with a strategic placement $P^*$, the effective distance achievable is at least 
\begin{equation}\label{eq:deff-tease}
    d_{\text{eff}}\geq \Big\lfloor \frac{d\times(2-\sqrt{1-f_e})+1}{2}\Big\rfloor - \epsilon d,
\end{equation} where $\epsilon d\approx -d/\log p<1$ for small codes. Indeed, when $\epsilon k<1$, Equation~\ref{eq:deff-tease} would predict $d_\text{eff}\geq \lfloor \frac{d+1}{2}\rfloor$ for surface code with standard qubits $d_\text{eff} \geq \lfloor \frac{2d+1}{2}\rfloor = d$ for a surface code with erasure qubits, as expected.

To optimize the placement strategy $P^*$, we prioritize achieving the optimal effective distance and then greedily place the remaining qubits to maximize impact on logical error rate. For a given number of erasure qubits, the optimal effective distance is achieved by placing the erasure qubits in $k$ full rows and columns, starting from the center, for the largest $k$ possible (using $2kd-k^2$ qubits). We will prove in Section~\ref{ssec:deff} that this guarantees the effective distance given in Equation~\ref{eq:deff-tease}. The remaining erasures are placed greedily to minimize the average Manhattan distance from the lattice center. See Figure~\ref{fig:placements} for an example. The greedy heuristic follows from an error path counting argument detailed in Section~\ref{ssec:paths}.

\subsection{The repetition code: a toy model}\label{ssec:rep-code}
\begin{figure}
    \centering
    \includegraphics[width=0.95\linewidth]{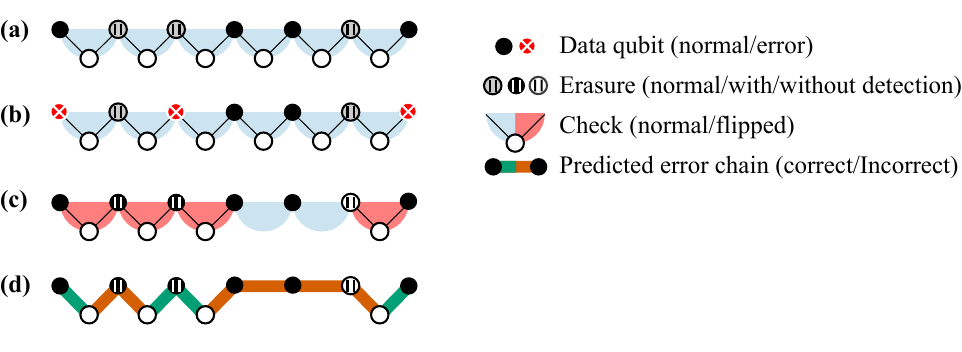}
    \caption{Error correction on the repetition code with heralded erasure noise. (a) We consider a distance-$7$ repetition code with $k=3$ erasure qubits, at indices $q_2,q_3,q_6$. (b) Suppose we encounter $3$ physical $X$ errors, triggering a detection event on $q_3$. (c) A physical error on $q_2$ may trigger the detection event, but does not flip any syndromes; this could be due to the deplorizing error being projected onto the code space as a $Z$ error or an identity, each with probability $25\%$. (d) There are two ways to match the syndromes. Since there's no detection even on $q_6$, we know the error must be the error chain not containing $q_6$.  }
    \label{fig:rep-code}
\end{figure}

We will first examine a repetition code as a toy example. Suppose our repetition code has $d$ data qubits with physical error rate $p$, $k$ of which are designated as erasures, and we are only interested in protecting against bit-flip ($X$-type) errors. When a heralded erasure occurs, the erasure qubit is reinitialized as a maximally mixed state, which is equivalent to applying a fully depolarizing error. After measuring the $Z$ observables, the heralded erasure error channel applies $I$, $X$, $Y$, or $Z$ with $1/4$ probability each, but only $X$ and $Y$ contribute to bit-flip errors. A maximum likelihood decoder \cite{dennis_topological_2002} is used to determine the error chain. We will calculate the rate of a logical bit-flip of the repetition code assuming ideal syndrome extraction and assuming that each data qubit experiences a bit-flip (or a heralded erasure error if it is an erasure qubit) independently with probability $p$.

First, note that there are always \emph{two possible explanations} for any syndrome patterns, which partition the data qubits into two disjoint subsets. See Figure~\ref{fig:rep-code}(d) for an example. This is because, on a 1D chain, each flipped check can only be matched to the left or to the right, and fixing the direction for any one check fixes the entire matching. 

Secondly, we observe that when $k>0$, a logical error occurs only when heralded erasure errors occur on all erasure qubits (otherwise, a maximum-likelihood decoder never makes mistakes, since the error chain cannot pass through the clean erasure qubits). This happens with probability $p^k$. It remains to find the logical error rate in this case. 

Suppose an error $E$ contains $l = l_d + l_e$ qubits ($l_d$ data qubits and $l_e$ erasure qubits). We have the following:

\begin{align*}
    &\prob[E] = p^{l_d} \times (1-p)^{d-k-l_d} \times (\frac12)^{l_e}\times (\frac12)^{k-l_e},\\
    &\implies \frac{\prob[E]}{\prob[E^c]} = (\frac{p}{1-p})^{2l_d-d+k}.
\end{align*}
A logical error occurs if $\prob[E]<\prob[E^c]$, or, $\prob[E]/\prob[E^c]< 1$, which implies:
\begin{align*}
        (2l_d-d+k)\log\frac{p}{1-p}< 0 \implies 
        l_d\geq  \lfloor\frac{d-k+1}{2}\rfloor.
\end{align*}
Hence, to leading order in $p$ and assuming $p\ll d$, the probability of a logical error given $E$ is 
\begin{align*}
    p_L &= p^k\times \prob[l_d\geq \lfloor \frac{d-k+1}{2} \rfloor ] \\
    &\leq p^k \times \frac{d+k}{2} \times p^{\lfloor \frac{d-k+1}{2} \rfloor}\\
    &\lesssim  p^{k + \lfloor \frac{d-k+1}{2} \rfloor} \\
    &= p^{ \lfloor \frac{d+k+1}{2} \rfloor},
\end{align*}
We can verify that the edge cases $k = 0$ and $k=d$ give the correct effective distances $d_\text{eff}$.

We can find $p_L$ exactly by summing over $E$:

\begin{align*}
    p_L = \prob_{E}[ \prob[E]< \prob [E^c]] = &\sum_{l_d\in [d-k]}\sum_{l_e\in[k]} {d-k \choose l_d}{k\choose l_e} 
    \\
    &\times  p^{l_d}(1-p)^{d-k-l_d}(\frac12)^{k}\\
    &\times \chi[2l_d< d - k]
\end{align*}
where $\chi$ is the indication function. We have verified that the approximation remains sufficiently accurate to the exact value for larger distances.

\subsection{Impact of erasure placement on logical error rate}\label{ssec:paths}
Unlike in 1D, where ways to match a syndrome pattern are always unique, there may exist exponentially many explanations for a syndrome pattern on a 2D surface code, making it challenging to generalize code-capacity level calculations. We call these the lattice traversing paths. In this section, we study the behaviors of small surface codes by enumerating all lattice traversing paths, and show that a logical failure can be approximately decomposed into logical errors along lattice traversing paths. This allows us to infer the correlation between an erasure qubit's placement and its impact on the logical error rate, justifies the row-and-column placement strategy, and allows us to carry out capacity-level analysis for effective distances in surface codes, as detailed in Section~\ref{ssec:deff}.

\emph{A lattice traversing path} is a sequence of 2D-connected nodes that starts and ends at two boundaries of a (possibly rectangular) lattice. The number of such paths across an $n\times m$ lattice, $P_{m,n}$, can be calculated with a recursive formula in polynomial time, and is often known as the number of unique ways a king can cross a $n\times m$ chessboard \cite{oeis_foundation_inc_-line_2024}. 

We can interpret logical errors on a surface code in a dynamic view as a king making its way across a chessboard. Indeed, a logical error occurs on a surface code only when there is a connected chain of physical errors connecting opposite boundaries; we can upper-bound the overall logical error using a union bound across logical errors corresponding to lattice traversing paths, and consider the effect of erasure allocation on each lattice traversing path separately. We will only include minimum-length paths in our analysis, since we are most interested in leading-order effects on the logical error rates. 

\begin{figure}
    \centering
    \includegraphics[width=0.5\linewidth]{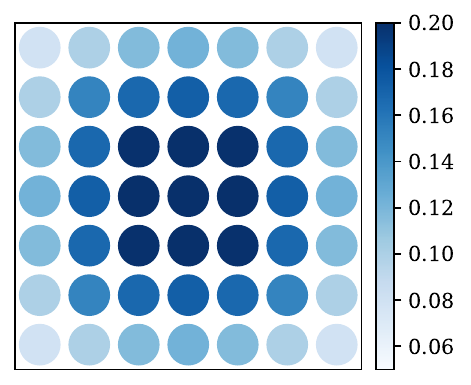}
    \caption{The fraction of error strings ($X$ or $Z$) that contain each physical qubit in a $d=7$ surface code.}
    \label{fig:theory-corr}
\end{figure}

Restricted to a minimum-weight lattice traversing path, logical errors on a lattice traversing path resemble those of the simple repetition code, where the impacts of erasure qubits on logical error rate and effective distance are more readily understood. This simplifying assumption allows us to quantify the influence of an erasure qubit by studying its importance on an ensemble of isolated lattice-crosing paths, suggesting a simple recipe for determining the optimal placement of an erasure qubit: if we can determine the benefit an erasure configuration $P$ brings to each lattice crossing path, we can bound its impacts on all logical errors to leading order. While this is difficult for arbitrary placements, we point out a few observations by studying lattice traversing paths on small codes that lead to our optimized placement strategy. 

We show in Figure~\ref{fig:theory-corr} a histogram of how frequently each qubit in a distance $7$ surface code is contained in a logical error. Immediately, we observe that the qubits in the middle are included in up to $3\times$ more paths than on the edge. We derive our closest-to-center placement strategy from this observation. Secondly, a more subtle point is that, since each lattice path either must traverse all columns or all rows, filling an additional row and column with erasures necessarily guarantees that each lattice path encounters an additional erasure qubit. The effective distance of the code will be guaranteed to increase by the same analysis given in Section~\ref{ssec:rep-code}. It is important to note that, while in general we should place erasures where they encounter more traversing paths (in the center), we should not prioritize this over filling additional columns/rows, since, to leading order, the logical error performance is constricted by the error with the shortest fault distance. 


\subsection{Surface code effective distance with hybrid-erasures}\label{ssec:deff}

Finally, we can find the total logical error rate of the surface code on a hybrid-$\mathcal{A}(d, f_e, P)$ architecture by decomposing a logical error into lattice traversing paths. Let $k$ be the largest integer such that $2kd-k^2\leq f_ed^2$. Then, we can place the erasure such that at least $k$ rows and columns are completely filled, such that any lattice traversing paths must cross at least $k$ erasure qubits. 

Let $E$ be a set of physical errors and let the recovery operator calculated by the decoder be denoted as $Dec(E)$. Then, a logical error occurs when $E \oplus Dec(E) = P$ for some logical operator $P\in\mathcal{P}_L$. Focusing on only the minimum weight logical errors to obtain an estimation to first order in $p$, we have:

\begin{align}
    p_L &=\prob_{E}\big[\bigcup_{P\in \mathcal{P}} E + Dec(E) = P \big]\\
    &\leq \sum_{P\in \mathcal{P}}\sum_{E\in P}\prob[E|P]\times   \prob_{E}[\prob[E]<\prob[P-E]]\label{eq:pl-surface}
\end{align}

where notice that regardless of $P$, $\sum_{E\in P}\prob[E|P]\times  \prob_{E}[\prob[E]<\prob[P-E]]$ describes exactly the logical error rate of a distance $d$ repetition code with $k$ erasures. Hence, Equation~\ref{eq:pl-surface} allows us to upperbound $p_L$ of a surface code by counting all lattice paths $\mathcal{P}$.

We can now arrive at the approximation of Equation \ref{eq:deff-tease}:
\begin{align*}
p_L &\lesssim 2^d\times p^{ \lfloor \frac{d+k+1}{2} \rfloor}\\
\implies d_{\text{eff}} &\gtrsim { \lfloor \frac{d+k+1}{2} \rfloor} + d\log_p2\\
&\geq \Big\lfloor \frac{d\times(2-\sqrt{1-f_e})+1}{2}\Big\rfloor - \epsilon d,
\end{align*}
where $\epsilon = -d/\log_{2}{p}$, and the last line follows from solving the quadratic equation $2kd-k^2\leq f_ed^2\leq 2(k+1)d-(k+1)^2$.

\section{Circuit-level evaluation methodology}\label{sec:eval-method}

While our analysis in a code capacity model sets expectations of the logical performance for varying erasure fractions and erasure placements, performance in a real device is better predicted via a circuit-level model. This captures error propagation, time-like differentiation of errors, and ancilla errors that arise due to simulating the underlying error-prone stabilizer circuit with a specific CNOT schedule. 

\subsection{Noise model}

To evaluate QEC performance, it is convenient to define a noise model that depends only on a single parameter $p$. This assumes that, as hardware fidelities continue to improve, the relative error rates of different error mechanisms remain the same. Table \ref{tab:noise-model} shows the noise parameters used in our evaluations, unless otherwise stated.

The transmon qubit noise model is a standard model in which initialization, readout, and two-qubit gates have error rate $p$, and single-qubit gates have a lower error rate $p/10$. To convert to a dual-rail error model, we consider the number of transmons involved in each operation. A dual-rail initialization or readout must be performed correctly on two transmons simultaneously, so these error rates double to $2p$. A dual-rail single-qubit gate is in fact a two-transmon gate, so this error rate becomes $p$. A dual-rail two-qubit gate only involves one transmon from each dual-rail qubit \cite{kubica_erasure_2023}, so this error rate remains $p$. Finally, erasure checks must be performed on the dual-rail qubit. Although current experimental demonstrations of this operation are relatively error-prone \cite{levine_demonstrating_2024}, we assume that the error can be reduced in the future to be on par with other operations, to an error rate of $p$.

We assume that we can perform a two-qubit gate between a dual-rail qubit and a standard transmon qubit. We believe this is not unreasonable, as transmon control is quite flexible and there is no fundamental barrier to such an operation. We assume that such a gate would also have error rate $p$, and an error in the gate would simultaneously erase the dual-rail and fully depolarize the transmon.

\begin{table}
\begin{tabular}{l|c|c}
    Error mechanism & Standard qubit & Dual-rail erasure qubit \\
    \hline
    Initialization error & $p$ & $2p$ \\
    Readout error & $p$ & $2p$ \\
    Single-qubit (H) gate error & $p/10$ & $p$ \\
    Two-qubit (CX) gate error & $p$ & $p$ \\
    Erasure check error & - & $p$ \\
\end{tabular}\\
\caption{Single-parameter noise model used in evaluations}
\label{tab:noise-model}
\end{table}

\begin{figure}
    \centering
    \includegraphics[width=0.7\linewidth]{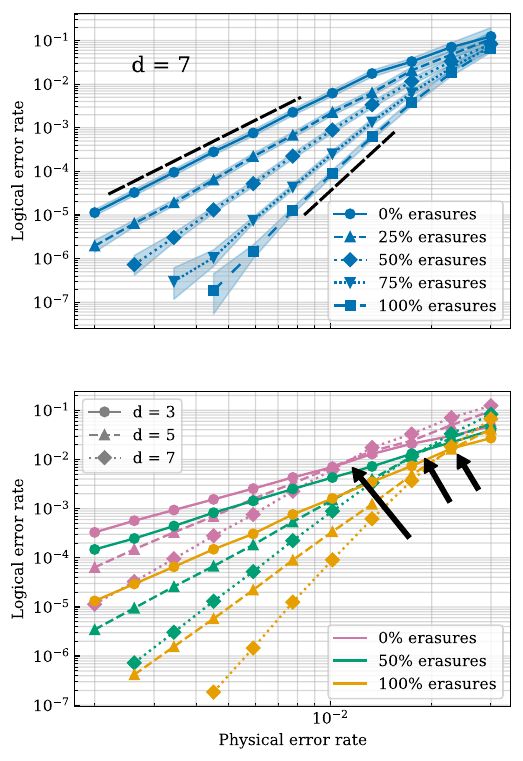}
    \caption{Logical memory performance of surface code of varying distance and varying fraction of erasure qubits, distributed according to our optimized heuristic. \emph{Top:} For fixed code distance $d=7$, increasing the fraction of erasure qubits improves $d_\text{eff}$, evidenced by a steeper slope. \emph{Bottom:} Increasing the fraction of erasure qubits also increases $p_\text{th}$, the crossing point for different distances.}
    \label{fig:lers}
\end{figure}

\begin{figure}
    \centering
    \includegraphics[width=0.9\linewidth]{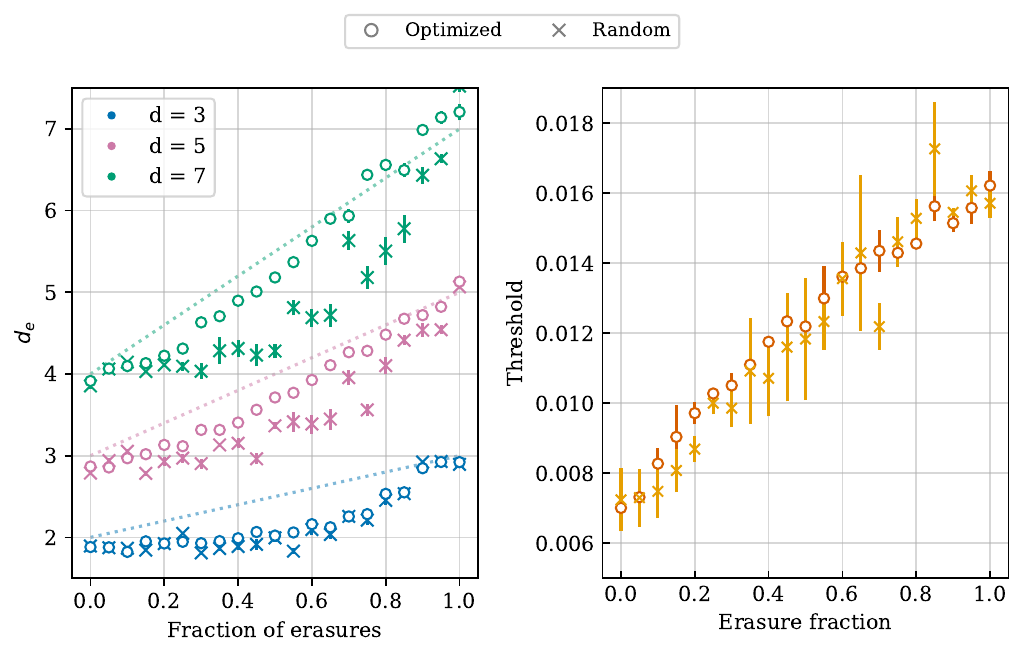}
    \caption{Error suppression characteristics for hybrid erasure architectures via circuit level simulation. \emph{Left}: Effective distance $d_\text{eff}$ of surface codes at various erasure qubit fractions. Optimized placement yields a higher effective distance for intermediate erasure fractions. Lines show linear interpolation between $(d+1)/2$ and $d$. \emph{Right}: Extracted surface code thresholds for random and optimized placement for different erasure fractions. Optimized placement leads to more consistent improvements in threshold.}
    \label{fig:d_eff}
\end{figure}

In this work, we assume that all erasure checks are perfect, such that there is never a false positive or false negative detection event. The impact of imperfect erasure checks has been explored in other works \cite{chang_surface_2024}, but accounting for imperfect erasure checks would significantly increase the complexity and computational cost of our simulations, so here we choose to simplify this aspect of the noise model so that we can focus on studying the interesting features of hybrid-erasure architectures without making additional assumptions on more noise parameters. Similarly, we consider the case where an erasure check is done after every time step in the circuit for simplicity, though the effects of different erasure check schedules have also been studied in previous work \cite{gu_optimizing_2024}.

\subsection{Simulation and decoding}

We perform circuit-level simulations using Stim\cite{gidney_stim_2021}. Errors are added upon initialization, readout, single-qubit gates, two-qubit gates, and erasure checks according to the noise model in Table~\ref{tab:noise-model}. One- and two-qubit gate errors on erasure qubits are modeled as heralded detections followed by a replacement of the qubit(s) with a maximally mixed state. The resulting circuit-level effect is an erasure flag along with $\{I,X,Y,Z\}$ errors with probability $\frac{1}{4}$ for each affected qubit. In Stim, the erasure flag is

Decoding is performed using PyMatching\cite{higgott_pymatching_2022}. To account for the additional knowledge from heralded errors on erasure qubits, the decoding graph is modified for each shot prior to decoding. Each gate error on an erasure qubit corresponds to a set of circuit-level errors $E$ correlated with a flag $f$. Depending on the sampled value of $f$, we make one of two modifications to the decoding graph. 1) If $f=0$, we remove the edges corresponding to the errors $E$ from the decoding graph. 2) If $f=1$, we update edge weights for each error $e \in E$ as follows:
\begin{align*}
    w_e = \log\Bigg[{\Bigg(\sum_{e'\in E}p_{e'}-p_e\Bigg)/p_e}\Bigg]
\end{align*}
where $p_e$ is the probability of error $e$ derived from the noise model in Table~\ref{tab:noise-model}. 

\subsection{Calculating effective distance and threshold}

To determine the effective distance and threshold of the code for a given experiment, we first manually identify an approximate threshold (where lines corresponding to different code distances cross in a physical vs. logical error rate plot) and consider only the points below the threshold. For the data corresponding to a particular code distance, we fit these points to the function $p_L = A (b\cdot x)^{d_\text{eff}}$ to find parameters $A, b,$ and $d_\text{eff}$. Then, we calculate the threshold $p_\text{th}$ as the average location where these fitted lines for different distances cross each other.


\subsection{Limitations}
In this work, we make several hardware and noise model assumptions that shape our findings. Firstly, our transmon cost scaling results depend on the assumption that an erasure qubit requires three transmons. In practice, transmon cost depends on erasure qubit implementation. In particular, if a cavity is used for readout as mentioned previously, the transmon cost of an erasure qubit would be 2 instead of 3. Secondly,  we assume in our noise model a specific ratio of erasure qubit error rates against standard qubit error rates (Table~\ref{tab:noise-model}), which our results are sensitive to. Quantitatively, reducing the cost of the erasure qubit or the ratio of erasure qubit-standard qubit error rates would contract the regions where we see hybrid architectures outperforming architectures with no erasure qubits or all erasure qubits -- however, qualitatively the analysis would remain the same, showing some parameter regions where the hybrid approach could achieve a target logical error rate with fewer transmons. 

More generally, the applicability of our results is limited by our choices in the noise model. As mentioned previously, we assume for simplicity that our erasure checks are spatially and temporally perfect, which is not hardware-realistic. Similarly, we assume that all errors in our erasure qubits are heralded erasures -- in reality, while heralded erasures are the dominant source of error in an erasure qubit, there is also some probability of unheralded Pauli errors. We leave further exploration of these and other changes to the noise model to future work. 

\section{Evaluating hybrid-erasure architectures}\label{sec:eval-result}

\begin{figure}
    \centering
    \includegraphics[width=\linewidth]{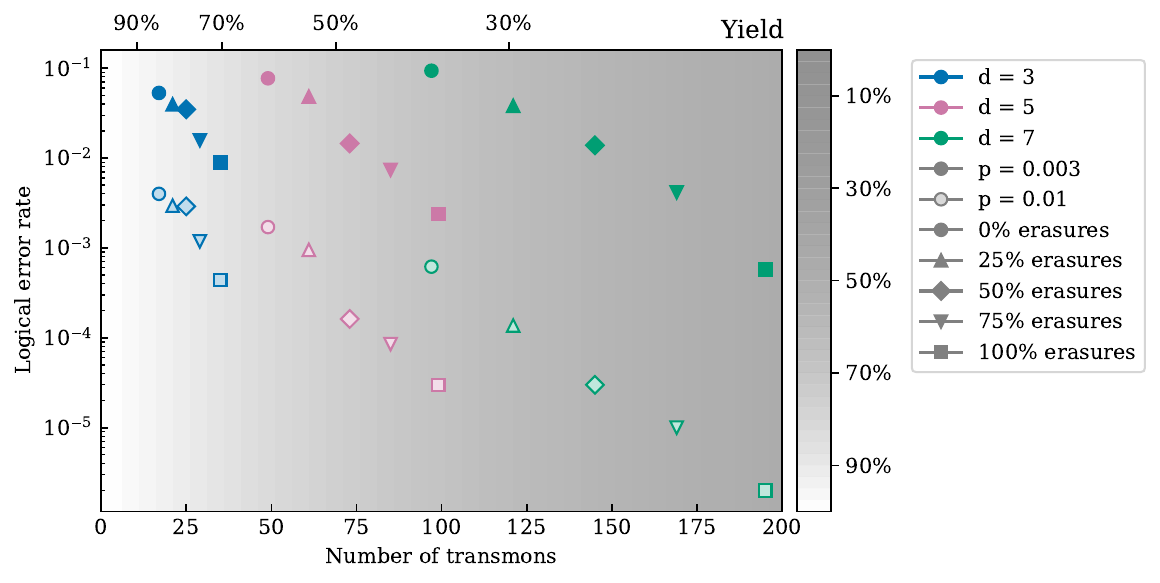}
    \caption{Achievable logical error rate for hybrid-erasure surface codes as a function of transmon count for several fixed physical error rates, where expected yield (background gradient) is calculated assuming a defect rate of $10^{-2}$ per transmon. Hybrid architectures outperform all-standard or all-erasure architectures for certain transmon budgets.}
    \label{fig:transmon_cost_yield}
\end{figure}

\subsection{Effective distance and threshold}

We first evaluate the logical error rate of surface codes with varying fractions of erasure qubits, distributed according to our optimized heuristic. The results are shown in Figure~\ref{fig:lers} and the key metrics of interest are the effective code distance $d_\text{eff}$ (the magnitude of the slope of each line) and the threshold error rate $p_{\text{th}}$ (the point where lines of the same erasure fraction cross). We observe both $d_\text{eff}$ and $p_{\text{th}}$ increasing as the fraction of erasure qubits increases, interpolating between the standard surface code and the expected values for a full erasure qubit architecture \cite{gu_optimizing_2024, chang_surface_2024, wu_erasure_2022}.

In Figure~\ref{fig:d_eff}, we plot the extracted effective distance fits for code distances 3, 5, and 7 at various erasure qubit fractions. We see a steady improvement in $d_\text{eff}$ as the erasure fraction increases. Additionally, we see that the optimized erasure qubit placement heuristic significantly outperforms random placement for intermediate erasure fractions, indicating the importance of erasure qubit location in hybrid architectures.

\begin{figure}[t]
    \centering
    \includegraphics[width=\linewidth]{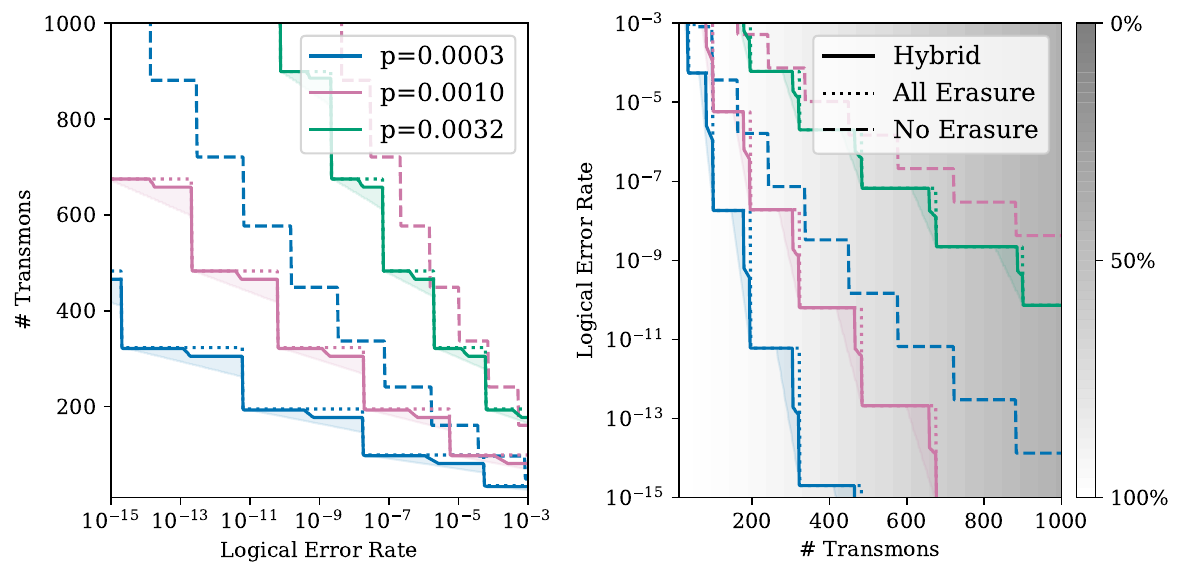}
    \caption{Scaling of costs in Figure~\ref{fig:transmon_cost_yield} for larger systems using an all erasure, no erasure, or hybrid erasure architectures. All hybrid points sweep over and select the best erasure fraction. Shaded region represents area of possible hybrid erasure values in between analytical upper bound and approximate observed circuit-level lower bound. \emph{Left:} Minimum possible logical error rate for fixed transmon cost. Background gradient indicates resulting single chip yield for a defect rate of $10^{-3}$. \emph{Right:} Minimum possible number of transmons for a target logical error rate.}
    \label{fig:scaling_analysis}
\end{figure}

\begin{figure*}
    \centering
    \includegraphics[width=\linewidth]{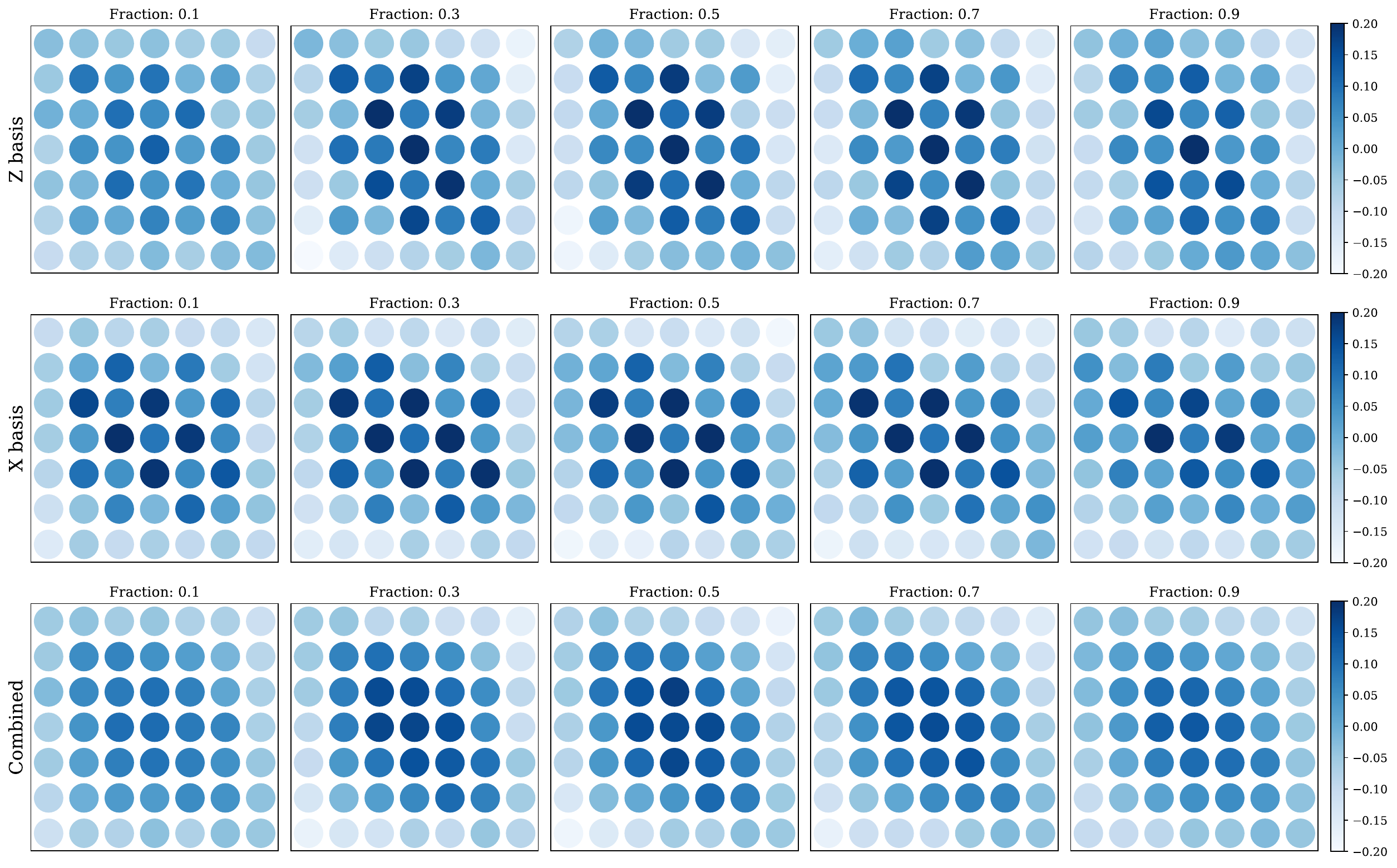}
    \caption{Effect of spatial placement of erasure qubits in a $d=7$ surface code. Color indicates correlation between logical error rate and presence of an erasure qubit at each location in the $7 \times 7$ grid of data qubits. Correlations are shown for $\bar Z$ (top), $\bar X$ (middle) and combined (bottom) error rates. ``Checkerboard'' patterns are evident in $\bar Z$ and $\bar X$ data but disappear for combined data.}
    \label{fig:spatial-corr-Z}
\end{figure*}

Figure~\ref{fig:thresholds} shows the extracted surface code thresholds for varying erasure qubit fraction, similarly showing a steady improvement as we add more erasure qubits. Again, a random placement of erasure qubits significantly underperforms compared to the optimized heuristic.

\subsection{Comparing transmon counts}

Dual-rail transmon qubits are particularly relevant when considering a hybrid-erasure architecture because they are built from the same components as standard transmon qubits. This makes a transmon-based hybrid-erasure architecture particularly compelling. While a typical transmon qubit consists of a single transmon, a dual-rail qubit instead requires three transmons \cite{gu_fault-tolerant_2023}. This is important because the yield is determined by the transmon count of a chip \cite{kreikebaum_improving_2020}, so reducing the transmon count can improve the scalability of manufacturing. Additionally, the number of control lines scales with the number of transmons, so heating concerns are also dependent on the transmon count rather than the qubit count.

In this context, we examine the transmon cost of hybrid-erasure architectures in Figure~\ref{fig:transmon_cost_yield} for several fixed physical error rates. These can be viewed as vertical slices through Figure~\ref{fig:lers}, where the cost $C(\mathcal{A}(d, f_e)) = d^2\times [(1-f_e) +3\times f_e]$ calculates the number of transmons needed to implement a surface code of distance $d$ and erasure fraction $f$. For a fixed code distance and physical error rate, we see the expected increase in transmon count and improvement in logical error rate as erasure fraction increases. For certain fixed transmon budgets, e.g. from 70 to 100 or from 150 to 190 transmons, a hybrid-erasure design gives a better logical error rate than can be achieved with a full-erasure design.

For a fixed chance of any transmon being defective (the defect rate $\epsilon$), we can translate transmon count $n$ into chip yield $y = (1-\epsilon)^n$, where we consider a chip to be operational only if no constituent transmons are defective. This effectively flips the x-axis, as shown in Figure~\ref{fig:transmon_cost_yield}. Again, we can that hybrid-erasure schemes allow us to interpolate between the standard qubit surface code and the erasure qubit surface code, allowing us to balance yield and logical error rate, as may be necessary to enable scalable modular quantum computers.

\subsection{Scaling to larger transmon counts}\label{ssec:scaling}

Although Figure~\ref{fig:transmon_cost_yield} helps analyze costs for near-term system sizes, we are also interested in the impact of hybrid erasure architectures for larger systems. To estimate the performance of the hybrid erasure architecture in higher distances and lower logical error rates, we use the empirical thresholds and a projected $d_\text{eff}$ bounded between two estimates. As an optimistic upper bound, we may expect the erasure fraction to behave as in the repetition code described in Section~\ref{ssec:rep-code}. We observe that the fit becomes more accurate as we increase the distance of the surface code. The analytical lower-bound is derived using Equation~\ref{eq:deff-tease}.  In Figure~\ref{fig:scaling_analysis}, we report the relative costs and performances of full-erasure, no-erasure, and hybrid architectures implementable over up to 1,000 transmons and down to logical error rates of $10^{-15}$. For each physical error rate, we estimate the logical error rates and the costs achieved by all possible erasure architectures, plotting the best logical error rates achieved by architectures implementable over fixed transmon budgets (\emph{Top}, Figure~\ref{fig:transmon_cost_yield}) and the fewest transmons used achieving fixed target logical error rates (\emph{Bottom}, Figure~\ref{fig:transmon_cost_yield}). The shaded regions represent the parameter regimes where a hybrid architecture may potentially outperform the standard approaches. We find that the best strategy is often to have $100\%$ erasure qubits, but there are ranges where the hybrid approach is potentially better. As the size of the system increases, these regions become proportionally smaller. We observe the same trend in the bottom plot of Figure~\ref{fig:scaling_analysis}.

We conclude that the benefits of a hybrid architecture diminish as the size of the system increases. However, in the near-term, there are potentially many opportunities where a hybrid architecture can improve logical-level performance. Furthermore, given the low yield of large chips, it may be preferable to keep the chips small. Indeed, in a fault-tolerant system built on concatenated codes~\cite{gidney_yoked_2023, pattison_hierarchical_2023}, it may be sufficient to use smaller distance surface codes where the results of Figure~\ref{fig:transmon_cost_yield} and Figure~\ref{fig:scaling_analysis} indicate that hybrid erasure qubits are more impactful. We leave further analysis of hybrid erasure qubits in a concatenated scheme to future work as the system scale needed to support such codes is likely beyond what is achievable in the near-term.

\section{Empirical Analysis of Erasure Placements} \label{sec:empirical}
In this section, we will extend our analysis on the impact of erasure qubit placement to the circuit level and compare against the code-capacity level predictions.

\subsection{Empirical effect of erasure qubit placement}
Figure~\ref{fig:spatial-corr-Z} shows logical error rate correlations for each qubit location in a $d=7$ surface code. Samples are generated from circuit-level simulations of randomly distributed erasure qubits of a given erasure fraction. As expected, we generally see that qubits near the center have the highest correlations with the logical error rate, since these qubits are included in the most error paths that connect opposite boundaries. Interestingly, we observe ``checkerboard'' patterns for independent $Z$ and $X$ logical error rates; we hypothesize that these are an artifact of the different CNOT orderings for $Z$/$X$ stabilizers. 
Combining the $Z$ and $X$ error rates into a single logical error rate recovers the expected behavior from the code-capacity analysis, where distance from the center is the most important feature.




\begin{figure}
    \centering
    \includegraphics[width=0.9\linewidth]{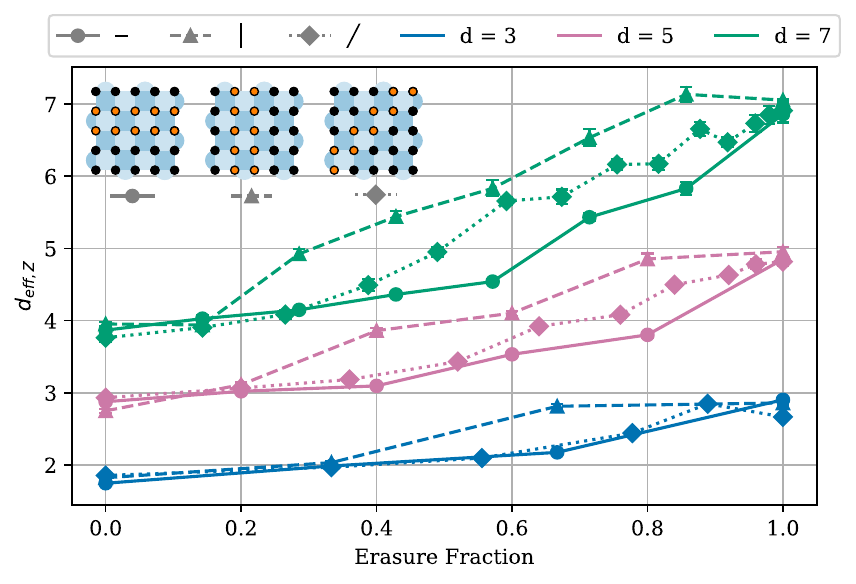}
    \caption{Placing erasure qubits in a way that intercepts as many traversing paths as possible maximizes effective distance gains in partial erasure architectures. Focusing on logical Z error paths, we see that adding a centrally-placed column of erasures, which encounters every possible logical Z error path, has a bigger effect on the effective distance than adding a diagonal, and an even bigger effect compared to adding a row.}
    \label{fig:deff_lines_comparison}
\end{figure}

\begin{figure}
    \centering
    \includegraphics[width=0.9\linewidth]{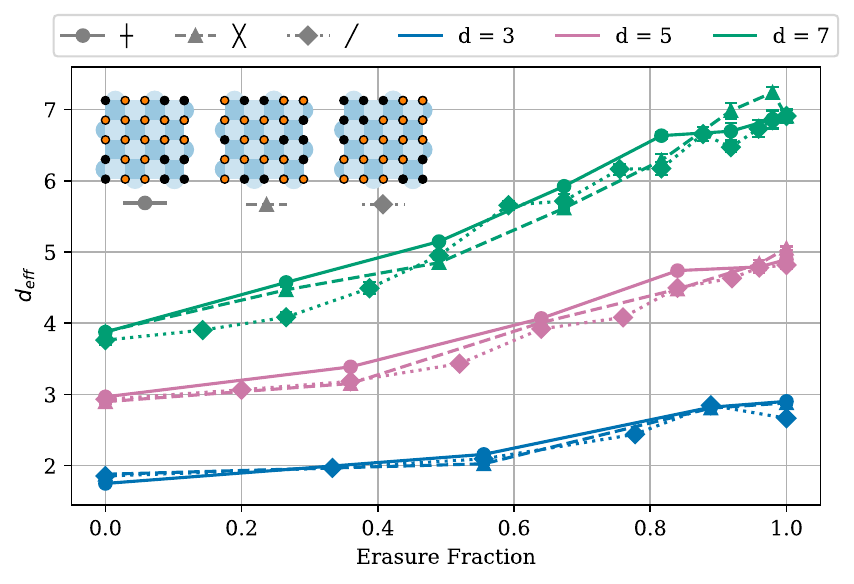}
    \caption{Placing lines of erasure qubits in different orientations has a small impact on the total logical error rate if the minimum number of erasures each logical error encounters is the same.}
    \label{fig:deff_cross_comparison}
\end{figure}

\begin{figure}
    \centering
    \includegraphics[width=0.7\linewidth]{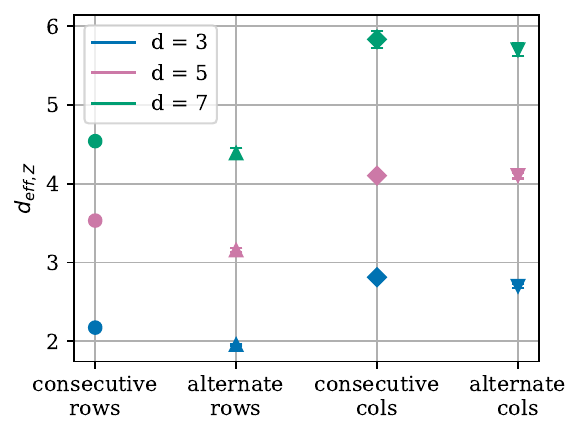}
    \caption{Gains in effective distance are slightly higher when lines of erasures are placed consecutively rather than alternately. Data shown is for $\frac{d+1}{2}$ lines of erasure, placed horizontally (rows) and vertically (cols).}
    \label{fig:deff_altvsconsec}
\end{figure}

\subsection{Other erasure configurations} \label{ssec:other_configs}
Finally, we investigate the performance of erasure placement with varying orientation. We compare the gains in effective distance achieved using each variation of erasure qubit placement, simulating each configuration at $d=3,5,7$ with increasing proportions of erasure qubits. Our results validate the heuristic priority to increase $k_e$, the minimum number of erasure qubits each logical error encounters. We show that the best effective distance can be obtained when the lines of erasures are placed in consecutive rows closest to the center as prescribed by the optimized placement strategy, while the orientations of  the configuration does not matter.

We first study the impact of the orientation of lines of erasures, demonstrating the importance of placing erasure qubits to maximize $k_e$. As shown in Figure \ref{fig:deff_lines_comparison}, we compare the cases where erasure qubits are organized in centrally placed rows, columns, and diagonals, focusing on the logical Z error rate. As we increase the number of lines of erasure qubits in each configuration, we see that the columns of erasure qubits show the greatest gains in effective distance, followed by the diagonals, and finally the rows. This is in keeping with our expectations: After adding a column, every logical Z error path crossing through the band of erasure qubits necessarily encounters an additional erasure qubit, increasing $k_e$ by one. In contrast, it takes adding 2 (or 3 in the corners) diagonals to guarantee an increase in $k_e$. Rows perform worst of all because they are parallel to the logical observable-- $k_e$ will be 0 unless every row is populated with erasure qubits, as there always exists an error path that does not encounter any erasure qubits.

Although our heuristic prioritizes increasing $k_e$, it does not prescribe a specific configuration if multiple can achieve the same $k_e$.  Finally, we will study the sensitivity of the logical error performance for varying orientation and centrality of lines of erasures.


First, we show that orientation of the lines do not matter as long as they achieve the same $k_e$. We show the $d_\text{eff}$ results obtained via simulation in \ref{fig:deff_cross_comparison} from three such configurations, where lines of erasures take different orientations, where the impact on effective distance is small.

Then, we see that it's better to place lines of earsures closest to the center. As shown in Figure \ref{fig:deff_altvsconsec}, we see that for an equal number of lines of erasures ($\frac{d+1}{2}$), placing them consecutively in the center yields a higher effective distance than placing them alternately. This follows from our results in Figures \ref{fig:theory-corr} and \ref{fig:spatial-corr-Z}, which show that we should expect the middle data qubits to have a higher impact on performance than those on the edge, as prescribed by the optimized placement strategy. 




\subsection{Validation and limits of code-capacity level analysis}\label{ssec:simulation-validation}

Although we did not expect capacity level analysis to yield accurate estimations of logical error rates, our goal was to gain insights about the performance of a hybrid erasure architecture and to capture key characteristics with simple models. With circuit-level simulation results in hand, we revisit the code-capacity analysis of Equation \ref{eq:deff-tease} and investigate the agreement between the simpler model and the empirical results. 

\begin{figure}
    \centering
    \includegraphics[width=0.9\linewidth]{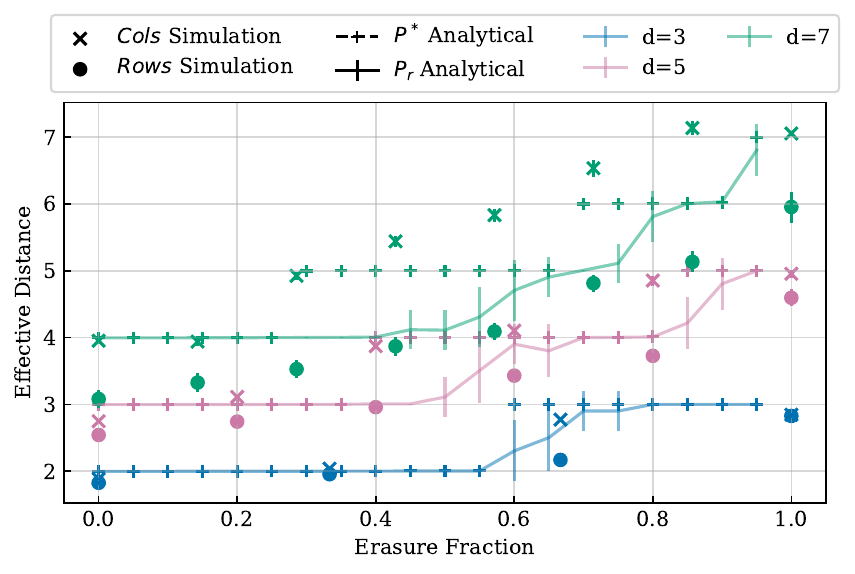}
    \includegraphics[width=0.9\linewidth]{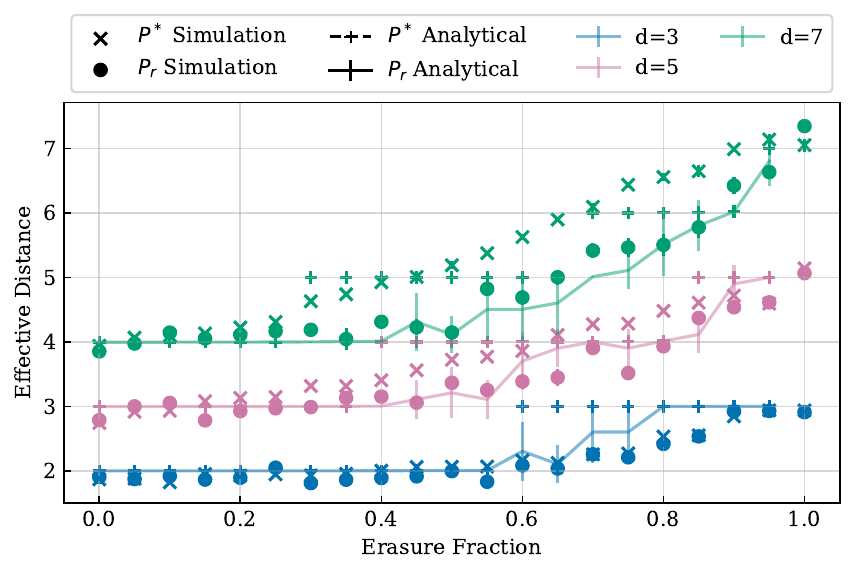}
    
    \caption{Comparing code-capacity model from Section~\ref{sec:hybrid-arch} with simulation data from Section~\ref{sec:eval-result} for two different placement strategies (random placement $P_r$, optimized heuristic $P^*$) and placements in consecutive rows and columns.}
    \label{fig:theory-vs-sim}
\end{figure}

First, we show in Figure~\ref{fig:theory-vs-sim} that the empirical effective distance $d_\text{eff}$ matches reasonably well between the predictions for effective blocking (consecutive columns) and placement strategies and $P^*$ (optimized), and between ineffective blocking (consecutive rows) and $P_r$ (random) across erasure qubit fractions for small distances. We have leveraged this analytical model in the projective cost studies of Section~\ref{ssec:scaling}, where it remains useful in parameter regimes where circuit-level simulations are computationally prohibitive. Secondly, the optimized placement strategy $P^*$, informed by insights at capacity level analysis, matches well between theory and circuit-level simulation. Finally, our importance metric based on lattice traversing paths can also be validated by the circuit-level correlations studied in Figure~\ref{fig:spatial-corr-Z}.

At the same time, deviations between the capacity-level analysis and circuit-level simulations highlight intriguing future directions for erasure qubits. For one, the channel-specific checkerboard pattern, only seen in the circuit-level correlation study, is likely an artifact of the syndrome extraction schedule designed to minimize hook errors. These variations suggest that a strategic placement of erasure checks may potentially help to identify and prevent error propagation, or to achieve asymmetric protection against different error channels. Secondly, rather surprisingly, the estimated effective distance from circuit-level simulation often outperforms capacity-level predictions. One possible explanation is that a decoder with circuit-level noise has more granular information about the space-time location of a heralded erasure error, which is lost on a phenomenological model that does not explicitly model syndrome extractions. This suggests that a closer inspection of space-time decoders designed for partial-erasure architectures may reveal additional performance gains. 

\section{Discussion}\label{sec:discussion}

Our work has explored hybrid erasure surface codes in which some fraction of physical qubits have an erasure bias and some do not. By combining theoretical code-capacity insights with detailed circuit-level simulations, we have shown that strategic placement of erasure qubits can improve both effective distance and threshold, opening up new avenues for practical error-corrected quantum computing. 

We have developed a well-motivated heuristic to place a limited number of erasure qubits in a larger surface code patch, prioritizing full rows and columns of erasure qubits. Using this heuristic, our circuit-level simulation of the hybrid erasure architecture validated the code-capacity analysis, showing a steady improvement in effective distance and threshold as the fraction of erasure qubits in the chip increases. With a noise model tailored to dual-rail qubits in superconducting transmon architectures, we found that a hybrid chip can achieve better logical error rates than homogeneous all-normal or all-erasure qubit chips for certain transmon-count budgets.

While our specific circuit-level noise model and transmon-counting results are targeted towards dual-rail superconducting implementations, our theoretical analysis and circuit-level simulation results pertaining to effective distance and spatial placement are broadly applicable to all types of erasure qubits.

We envision several intriguing directions for future work. In this study, we assumed perfect erasure checks (no false positives or negatives), but the performance of erasure qubits can heavily depend on the accuracy of the checks \cite{chang_surface_2024}; in addition, there are several aspects of erasure qubits that can be optimized or balanced, such as the frequency of erasure checks \cite{gu_optimizing_2024}. It would also be interesting to study the performance of hybrid-erasure schemes in the presence of defects, as has been done for the standard surface code \cite{auger_fault-tolerant_2018, debroy_luci_2024}. Finally, we are interested in exploring how a hybrid-erasure chip performs under biased noise by adding more rows or columns of erasure qubits as an alternative to nonsquare or XZZX surface codes. Erasure qubits placed along rows or columns may be an effective way to protect one logical observable more than the other if there is a bias in physical noise.

Overall, our work demonstrates that a hybrid-erasure architecture is a viable pathway toward enhancing the logical performance of surface codes while efficiently managing hardware resources. These results lay a strong foundation for future studies to address imperfect erasure checks, biased noise, and defect tolerance, and they provide an exciting roadmap for adapting our approach to various quantum hardware platforms.

\section*{Author contributions}
The authors devised the project idea together and all helped guide its direction. W.Y. performed theoretical analysis and processed spatial correlation data. J.D.C. and J.V. developed circuit-level simulation code. J.D.C. and M.H.T. performed circuit-level simulations. J.V. performed scaling cost analysis. All authors wrote and edited the manuscript and figures.

\section*{Acknowledgements}

We are grateful to Connor Hann for insightful discussions on partial-erasure architectures and for motivating us to pick up this project again after a long hiatus.

This work is funded in part by the STAQ project under award NSF Phy-232580; in part by the US Department of Energy Office of Advanced Scientific Computing Research, Accelerated 
Research for Quantum Computing Program; and in part by the NSF Quantum Leap Challenge Institute for Hybrid Quantum Architectures and Networks (NSF Award 2016136), in part based upon work supported by the U.S. Department of Energy, Office of Science, National Quantum 
Information Science Research Centers, and in part by the Army Research Office under Grant Number W911NF-23-1-0077. The views and conclusions contained in this document are those of the authors and should not be interpreted as representing the official policies, either expressed or implied, of the U.S. Government. The U.S. Government is authorized to reproduce and distribute reprints for Government purposes notwithstanding any copyright notation herein. This work was completed in part with resources provided by the Research Computing Center of the University of Chicago.

FTC is the Chief Scientist for Quantum Software at Infleqtion and an advisor to Quantum Circuits, Inc.

\bibliographystyle{IEEEtran}
\bibliography{references}

\end{document}